\documentclass[]{elsart}

\usepackage{amsfonts}
\usepackage{amsmath}
\usepackage{amssymb}
\usepackage{graphicx}

\journal{JOURNAL OF ALLOYS AND COMPOUNDS}
\begin{document}
\begin{frontmatter}



\title{Molecular dynamics study on planar clustering of xenon in UO$_{2}$}


\author[cor1,tu]{H. Y. Geng},
\author[tu]{Y. Chen},
\author[tu]{Y. Kaneta},
\author[mt1,mt2]{M. Kinoshita}
\corauth[cor1]{Corresponding author.}
\address[tu]{Department of Quantum Engineering and Systems Science, School of Engineering, The University of Tokyo, Tokyo 113-8656, Japan}
\address[mt1]{Nuclear Technology Research Laboratory, Central Research Institute of Electric Power Industry, Tokyo 201-8511, Japan}
\address[mt2]{Japan Atomic Energy Agency, Ibaraki 319-1195, Japan}
\begin{abstract}
Interatomic potentials of uranium dioxide
are investigated on their applicability to model structural stabilities
beyond fluorite phase by comparing with \emph{ab-initio}
results. A high pressure cotunnite phase and loosely stacking virtual crystal are
involved in order to get a primary confidence for large-scale deformation modelings of
UO$_{2}$ under irradiation damages. The behavior of Xe atoms
in UO$_{2}$ fuel is studied with molecular dynamics simulations.
Besides the ground state bubble geometry, we find that a planar distribution
is also (meta-)stable for xenon under thermodynamic perturbations. The Xe atoms with a planar configuration are in a liquid state
at a typical reactor temperature
of 1000\,K, which presents a modulated layer-structure near the interface with solid
UO$_{2}$. A planar defects loop remains after these Xe atoms are released out.
\end{abstract}

\begin{keyword}
Uranium dioxide \sep Molecular dynamics \sep Planar defects \sep Solid-liquid interface
\PACS 34.20.Cf \sep 71.15.Nc \sep 61.72.Nn \sep 71.15.Pd \sep 68.08.-p
\end{keyword}
\end{frontmatter}

\section{Introduction}
\label{sec-Intro}
Uranium dioxide (UO$_{2}$) is a standard nuclear fuel in pressurized-water reactors.
In operation in power plants or in the context of direct disposal of spent fuel,
a clear understanding of its structural, thermodynamical and kinetical properties
is very important.
Therefore many studies have been undertaken about the behavior of this material
under irradiation. The crystallographic, elastic and thermodynamic properties,
static and high frequency dielectric constants, point defects formation energies,
activation energies for lattice migration have been studied very well today and can be
easily found in abundant literature \cite{matzke87,jnm,jap,ronchi94,lay70,hutchings87}.
A number of work on molecular dynamics (MD) simulations for UO$_{2}$ also has been
published in recent years with classical interatomic potentials to get various thermodynamic
and transport properties \cite{meis05,basak03,morelon03,yamada00-1,yamada00-2}.
However, most of these studies focused on the stable fluorite
phase, and bears a limitation that actual UO$_{2}$ might deviate
from this structure locally due to irradiation damages.
It is the case especially in cascade
simulations \cite{brutzel03,delaye05}. This provokes
a question about the applicability of classical potentials to model the behavior of
UO$_{2}$ with lager-scale deformations with defects.

On the other hand, as a kind of major fission product, the behavior of xenon (Xe) atoms in
uranium dioxide also has attracted considerable experimental and theoretical
attentions. Concern has been focused particularly on its important role in fuel
swelling and this has accordingly led to a desire to obtain a greater understanding
of the basic processes governing its migration and trapping within the fuel.
The behavior of single isolated atoms of Xe in stoichiometric UO$_{2}$ and its
interaction with Schottky vacancies have been well studied \cite{jackson85,jackson85-2,macinnes82,nicoll95}. In this paper we prefer
to investigate a large-scale collective behavior of Xe atoms.

From a point of view of experiment, as a kind of nuclear fuel, uranium dioxide has extensive sub-grain
formation (grain sub-division) under energetic fission fragment
irradiations. The restructuring starts after accumulating certain
amount of rare gas as fission products, such as xenon in the matrix.
Observing experimental results, one can assume that the process of sub-grain formation is
initiated by bubble or planar structure of nanometer scale through
decomposition and reformation of U-O-Xe complex.
Sub-grains develop from a repeated
cycle of damage (ballistic excitation and local heating), recovery
(relaxation) and quenching. To investigate this dynamical process,
molecular dynamics (MD) study will be employed using interatomic potentials
with long range Coulombic interactions. In this paper,
we show that a planar defect on [111] plane in UO$_{2}$ fluorite
structure seems (meta-)stable when a nanometer scale of the defect is
created. It is reasonable because the relaxation to the ground state bubble
geometry might be prevent by an energy barrier originated from the rearrangement of local atomic
environment. The mechanism, which we find by MD simulations, could be
useful for understanding the nuclear fuel
restructuring process at high burnup irradiations.

In following sections, we compare the results of empirical shell model and pair
potential model with first-principles calculations. The variation of relative energies and structural stabilities
of different phases of UO$_{2}$ is studied, to examine
the transferability of these potentials beyond the standard fluorite phase. In section \ref{sec:MD}, we
will discuss the molecular
dynamics simulations of a planar clustering of xenon on [111] plane of fluorite UO$_{2}$ in details,
followed by a conclusion.

\begin{table}
  \centering
  \caption{Parameters of pair potentials between Xe and UO$_{2}$}\label{tab:LJ-param}
  \begin{tabular}{l c c c c}
    \hline\hline
     & Xe-Xe (LJ) & Xe-O (LJ) & Xe-U (BM) & Xe-U (BM-fitted)\\
    \hline
    $\varepsilon$ (eV) & 0.017 & 0.0099 & 2787 &4887.7\\
    $\sigma$ (A) & 4.29 & 2.50 & 0.41426 &0.415 \\
    \hline\hline
  \end{tabular}
\end{table}

\section{Transferability of interatomic potential}
\label{sec:relativeE}
\subsection{Method of calculation}

Interatomic potentials are checked by calculating the energy variation of bulk materials with
density functional theory. Plane-wave pseudopotential method implemented in the
code VASP \cite{vasp,kresse96} is used. Pseudopotentials model the potential yielded by nuclei
and core electrons whereas the valence electron wave functions are expanded
in a plane-wave basis. Spin-polarized local density approximation (LSDA)
with Hubbard U term correction of the energy
functional (LSDA+U) \cite{anisimov91,anisimov93} is employed. The parameters
for Hubbard term are taken as $U=4.5$\,eV and $J=0.51$\,eV,
which have been checked carefully by S. L. Dudarev \emph{et al.} \cite{dudarev00,dudarev97,dudarev98}. All
calculations employ projector-augmented wave (PAW) pseudopotentials \cite{blochl94,kresse99} with a cutoff
of kinetic energy for plane waves of 400 eV. Integrations in reciprocal space are performed in
the first Brillouin zone with 18 irreducible k-points for fluorite structure and at least 28
irreducible k-points for other phases generated with the Monkhorst-Pack scheme \cite{monkhorst76}. The
convergence of this parameter set is well checked. The energy tolerance for the charge self-consistency convergence
is set to $1\times10^{-5}$\,eV for all calculations.

For the part of interatomic potential models, newly parameterized shell model by
C. Meis \cite{meis05} and pair potential model by C. B. Basak \cite{basak03} are used.
The former provides the most accurate static and thermodynamic description
for fluorite UO$_{2}$ whose details are listed in the table 1 of Ref.\cite{meis05},
and the latter also gives a reasonable result whose parameters are given in table 1
of Ref.\cite{basak03}. When xenon is presented, a short ranged pair potential model is adopted
to describe interactions between Xe atom and UO$_{2}$. A Lennard-Jones (LJ) model
\begin{equation}
  V(r_{ij})=\varepsilon_{ij}\left[\left(\frac{\sigma}{r_{ij}}\right)^{12}-\left(\frac{\sigma}{r_{ij}}\right)^{6}\right]
\label{eq:LJ}
\end{equation}
is employed for Xe-Xe and Xe-O pairs whose parameters are determined by fitting to first principles
calculated total energies of FCC-Xe and XeO$_{3}$ in CuAu$_{3}$ phase, respectively \cite{Kaneta}.
The Xe-U pair follows Jackson's adoption
\cite{jackson85} which is a Born-Meyer (BM) function
\begin{equation}
  V(r)=\varepsilon\cdot\exp\left(-\frac{r}{\sigma}\right).
\label{eq:BM}
\end{equation}
All parameters of these potentials are listed in table \ref{tab:LJ-param}.

It is necessary to point
out that the Xe-U potential parameterized by Jackson is as a part of a shell model where he studied
the Xe-UO$_{2}$ system \cite{jackson85}. Therefore the performance of this potential
may have some ambiguity. We improve it by fitting to \emph{ab initio} energies
and list its parameters in table \ref{tab:LJ-param} as BM-fitted.
To calculate the energy with classical potentials, GULP \cite{gale97} code is employed
where Ewald summation is involved to treat the long-ranged Coulomb interactions among
uranium and oxygen atoms.
To ensure the comparability of calculated energies, all structures have their geometry fixed both for \emph{ab initio}
and classical potential calculations. The cutoff radius of short-ranged Van der Waals
potentials are set to 10.0\,{\AA} and fixed for all calculations.

\subsection{Results and discussions}

To check the transferability of interatomic potentials beyond fluorite phase, cohesive energy in
cotunnite phase with a space group of $Pnma$ that stabilizes under
high pressure is calculated. An experiment showed the pressure-induced phase transition
from fluorite to cotunnite phase starts at $\sim$42\,GPa \cite{idiri04}, whereas a first-principles
calculated setout pressure is about 38\,GPa \cite{geng06}.
Local pressure due to irradiation effects is usually higher than this threshold value
and the excess energy transferred from irradiation tracks to lattice can overcome
the possible barrier easily. Therefore it is very possible that a portion of cotunnite phase of UO$_{2}$
appears in irradiation damaged fuel.

\begin{figure}
\centering
  \includegraphics*[0.22in,0.19in][3.9in,2.96in]{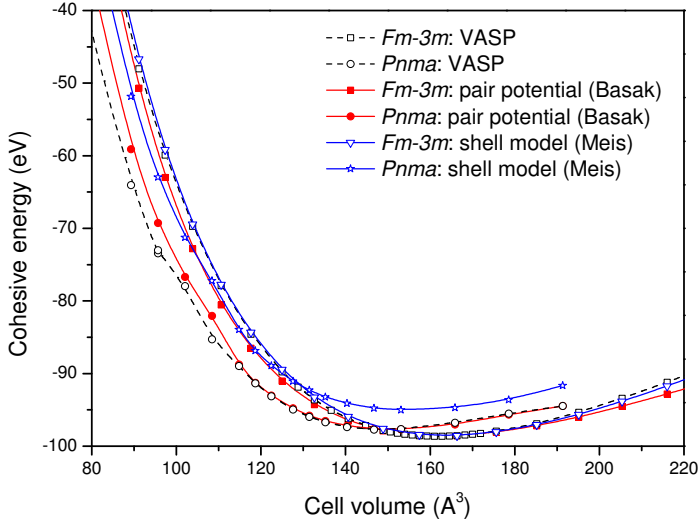}
  \caption{(Color online) Cohesive energies of UO$_{2}$ in fluorite ($Fm\overline{3}m$) and
  cotunnite ($Pnma$) phases calculated by \emph{ab initio} (VASP) compared with those
  predicted by interatomic potential models (shell model \cite{meis05}
  and pair potential \cite{basak03}). Note the latter curves have been shifted to have
  the same reference energy as \emph{ab initio} results.
\newline
}
  \label{fig:revE-vasp-shell-pp}
\end{figure}

The cell volume dependence of UO$_{2}$ cohesive energy is shown in figure \ref{fig:revE-vasp-shell-pp},
where each unit cell contains 12 atoms (U$_{4}$O$_{8}$). The cohesive energy of fluorite phase
at a volume of 166.375\,$\mathrm{\AA}^{3}$ calculated by VASP is taken as the reference point.
The energies predicted by interatomic potential models are shifted so as to have the
same value as \emph{ab initio} results. It is evident from figure \ref{fig:revE-vasp-shell-pp}
that in high-compression and stretching regime the pair potential (Basak) model becomes worse although
a reasonable variation of energy is predicted around the equilibrium volume neighborhood
compared with \emph{ab initio} results. On the other hand, the shell model (Meis) provides a rather
accurate energy variation for fluorite phase ($Fm\overline{3}m$) within the whole studied volume region.
However a big deviation is observed for cotunnite phase and the predicted
energy variation is much poor by comparison with pair potential model. In this
sense the transferability of shell model is unacceptable to treat UO$_{2}$ beyond fluorite structure.

\begin{figure}[!htb]
\centering
  \includegraphics*[0.22in,0.19in][3.9in,2.96in]{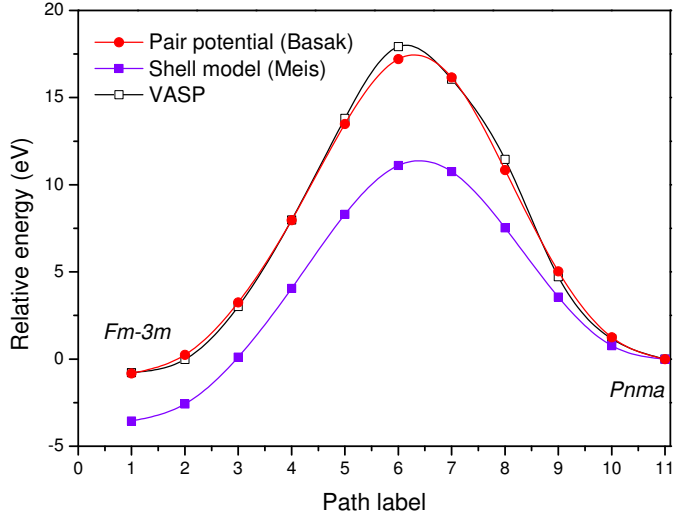}
  \caption{Comparison of the relative energy of UO$_{2}$ from $Pnma$ to $Fm\overline{3}m$ phase along
  a designed path calculated by VASP with interatomic potential models. The
  cohesive energy of $Pnma$ phase is set as the reference point.
}
  \label{fig:revE-pnma2fm-3m}
\end{figure}

A series of intermediate structures along one possible
transition path from $Fm\overline{3}m$ to $Pnma$ phase are also investigated\footnote{
This path is designed according to a principle of least displacement of atoms \cite{geng06}, which
is assumed to be the most possible candidate for the true transition path.
However, whether it is a true path or not does not affect the conclusion made here.}.
The variation of energy along this path is shown in figure \ref{fig:revE-pnma2fm-3m}. The energy of cotunnite phase has been
set as the reference point and all curves are shifted to have the same value at
this point. The shell model parameterized by Meis has some deviations,
the largest deviation from first-principles result occurs at the middlemost
phase where an energy difference of $\sim$6.8\,eV for 12 atoms is obtained.
The shell model has a large error in description structural phase transitions
of UO$_{2}$. Obviously this fault originates from the classical treatment of electric
polarization in the semi-empirical atomic model and damaged its transferability drastically.
It seems to be a congenital problem for shell model and we can expect a similar situation when apply it to other ionic materials.

\begin{figure}[!b]
\centering
  \includegraphics*[0.22in,0.19in][3.9in,2.96in]{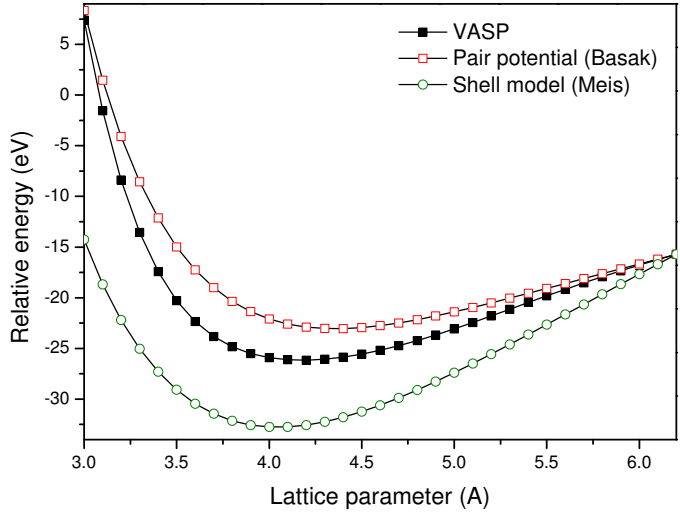}
  \caption{Variation of the relative energy of UO$_{2}$ in a cubic virtual structure (see text)
  along its lattice parameter predicted by VASP compared with interatomic potential models. The
  end points of these curves are set as the reference point.
}
  \label{fig:revE-vstuc}
\end{figure}

When Xe atoms congregate to a local region and form, for example, bubbles, the
structure of UO$_{2}$ near the interface boundary will departure from standard
fluorite phase greatly and become loosely stacked. To investigate the applicability
of interatomic potential models to this kind of circumstance, we employ a loose virtual structure
in which one uranium atom occupies at (0, 0, 0) and two oxygen
atoms occupy (1/4, 1/4, 1/4) and (3/4, 3/4, 3/4) sites in a cubic cell. Comparing
with fluorite phase, this virtual structure can be viewed as derived from it by removing
three uranium atoms on the face centers as well as the bonded oxygen atoms.
Based on above discussions, we can expect a failure of the shell model on this phase.
The variation of cohesive energy plotted in figure \ref{fig:revE-vstuc}
proves that it is true. The shell model parameterized by Meis \cite{meis05} predicted
a much deeper binding energy by comparing with \emph{ab initio} result.
The increment of energy under compression is slow while it is too fast under
stretching. Oppositely, the performance of pair potential model is acceptable
at least, although a shallower binding energy, a smaller lattice parameter and bulk modulus
are predicted.

\begin{figure}[!b]
\centering
  \includegraphics*[0.22in,0.19in][3.9in,2.96in]{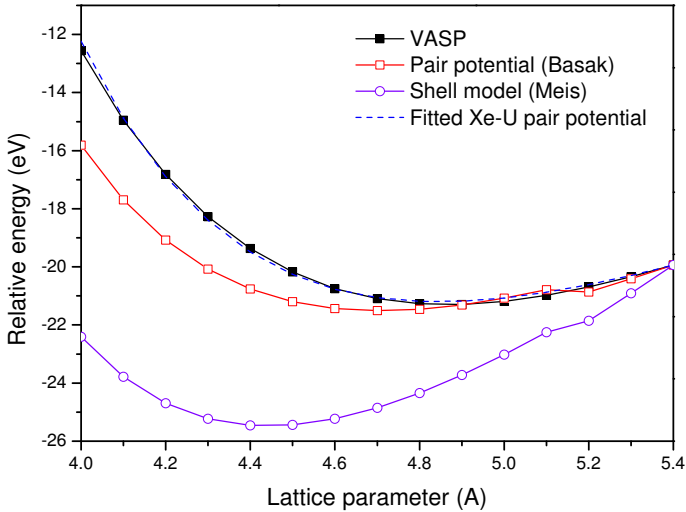}
  \caption{Variation of the relative energy of UO$_{2}$Xe in a cubic virtual structure (see text)
  along its lattice parameter predicted by VASP compared with interatomic potential models. The
  end points of these curves are set as the reference point. Pair
  potentials are used to describe Xe-UO$_{2}$ interactions (see table \ref{tab:LJ-param})
  and the dashed curve
  is the result fitting to the \emph{ab initio} curve to adjust the Xe-U interaction parameters (listed in table \ref{tab:LJ-param} as BM-fitted).
}
  \label{fig:revE-vstuc-xe}
\end{figure}

When one Xe atom is inserted into the (1/2, 1/2, 1/2) site of the previous loose virtual cell, as figure \ref{fig:revE-vstuc-xe}
showing, the performance of shell model deteriorated further. Its huge
deviation from \emph{ab initio} results bereaves its applicability
to this circumstance. Under high compression region, the pair potential
model also becomes poor, whose interaction of Xe-U pair is
taken after Jackson that labeled as BM in table \ref{tab:LJ-param}.
This potential was not well parameterized because it was for a
shell model \cite{jackson85}. To improve its quality in pair potential model,
we re-parameterize it by fitting to the \emph{ab initio} cohesive energy curve.
The resulted energy variation (associating with pair potentials of UO$_{2}$
parameterized by Basak) is shown as the dashed line in figure \ref{fig:revE-vstuc-xe}.
It is necessary to point out that this treatment is expedient to get qualitative understanding
of xenon behavior in UO$_{2}$.

\section{Classical molecular dynamics simulation}
\label{sec:MD}
\subsection{Simulation method}

To model the collective behavior of xenon in UO$_{2}$, the classical molecular
dynamics method implemented in DL\_POLY\_2 \cite{dlpoly} code is employed,
as well as the Basak's potential set for UO$_{2}$ \cite{basak03}.
When Xe atoms are involved, potentials listed in table \ref{tab:LJ-param}
is added to the Basak's set where the parameters of (BM-fitted) rather than (BM) are adopted for Xe-U pairs.
Previous section shows this set of potentials gives an acceptable description of
energy variation for UO$_{2}$-Xe system around large scale deformations, which provides us a confidence that
the following simulations are qualitatively correct leastwise.

It is well known that Xe does not dissolve in UO$_{2}$, which means phase separation
is inevitable in nuclear fuel when xenon is presented as a kind of fission
product. Usually this will lead to bubbles of Xe atoms. In nuclear reactor, however,
Xe atoms (or clusters) may have other meta-stable structures
around grain boundaries.
Since the behavior of bubbles has been studied extensively \cite{macinnes82,jackson85-2},
in this paper we prefer to focus mainly on a possible planar distribution of xenon in UO$_{2}$.

\begin{figure}[!b]
\centering
  \includegraphics*[scale=0.4]{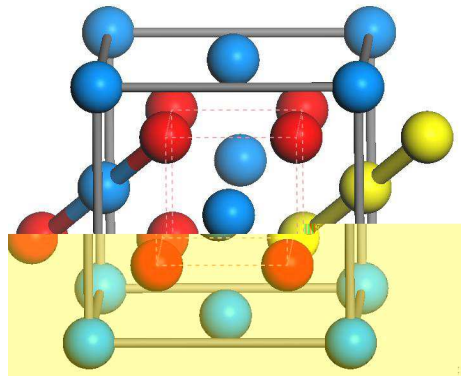}
  \caption{(Color online) The fluorite structure of UO$_{2}$, where the yellow trios
  along (111) direction on some [111] planes are partly substituted by Xe atoms randomly when
  preparing initial structures for MD simulation.
}
  \label{fig:fluorite}
\end{figure}

Previous studies had suggested that Xe atoms diffuses in trivacancies (that is, bound Schottky trios
consisting of a uranium vacancy and two oxygen vacancies) \cite{matzke66,matzke67,catlow78,nicoll95}.
On the other hand, first-principles calculation using pseudopotentials approach
revealed that a vacancy trio with a geometry lining up along (111)
direction is the most stable one \cite{Iwasawa06}. Therefore we intermingled
UO$_{2}$ matrix by replacing bonded UO$_{2}$ trios (the yellow balls in figure \ref{fig:fluorite})
with Xe atoms on [111] planes randomly when prepare the initial configuration for MD simulations. [111]
plane is chosen just because it has the largest layer interval distance
which can mostly preserve the UO$_{2}$ matrix when Xe is doped.

It is quite clear that the conventional cell unit of fluorite phase showed in
figure \ref{fig:fluorite} is inconvenient when simulating layer-related behavior
along (111) direction. A three times larger cell unit with (111) direction
aligning with Z axis is used instead. A $6\times6\times6$ supercell based on this
enlarged unit cell is employed for following MD simulations, corresponding to 648 conventional unit cells
of fluorite structure and more than 7000 atoms. To check the size effects, we also use some supercells of $12\times12\times6$ and
$6\times6\times12$, having nearly 30000 and 14000 atoms, respectively.
In all simulations, a NPT ensemble is employed with zero pressure and 1000\,K.
To remove the artificialness introduced when preparing initial geometries, all configurations
are heated up to 3000\,K and then cool down firstly. 5000
simulation steps are used to equilibrate system before gathering data.
The timestep is set as 1\,\emph{fs}; the cut-off radius is 7.5\,{\AA} for short-ranged
Van der Waals interactions and 12\,{\AA} for long-ranged Coulomb interaction
in real-space part of Ewald summation, with a Ewald precision of $1.0\times10^{-5}$.

\subsection{Results and discussions}
\begin{figure}[!b]
\centering
  \includegraphics*[0.22in,0.19in][4.0in,2.96in]{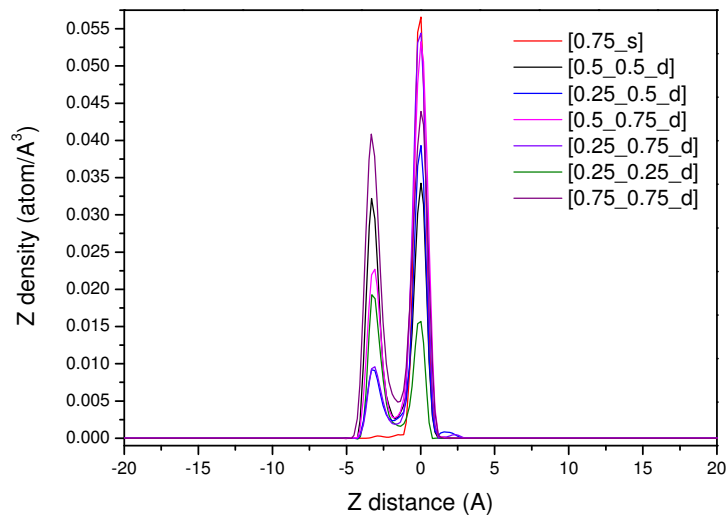}
  \caption{(Color online) Variation of number densities of Xe along (111) direction in
  UO$_{2}$ simulated at 1000\,K.
  Seven different kinds of initial distribution of Xe on [111] planes are considered.
  }
  \label{fig:Zden1}
\end{figure}

Although it is well known that bubble is the most stable geometry for Xe in UO$_{2}$, our following simulations show that if
the initial distribution of xenon is far from a bubble geometry and close to, for example, a planar
distribution, then the bubble structure can not be recovered just
by usual thermodynamic perturbations. It is perhaps due to a high energy barrier
between these two configurations since many atoms are involved for structure rearrangement.
To check the stability of our simulation result,
the dependence on number density of Xe is investigated. As figure \ref{fig:Zden1}
shows, seven initially different configurations with Xe distributed on [111] planes
give very similar results.
Here only the variations of number density of xenon along (111) direction (Z axis) are
plotted. The nomenclature of curves is as follows: decimal fractions denotes the
ratio of xenon atoms to the total heavy atoms (xenon and uranium) on the same
[111] plane, and the last letter indicates how many consecutive layers are
involved (namely, ``s'' stands for single layer and ``d'' for double layers). For example [0.25{\_}0.5{\_}d] means two conservative [111] layers
are doped, in which quarter uranium atoms on the first plane and half
of the second plane are randomly substituted by Xe. Figure \ref{fig:Zden1} illustrates
that a layer-like structure is stable for a range of density of Xe in UO$_{2}$.

\begin{figure}[!htb]
\centering
  \includegraphics*[scale=0.5]{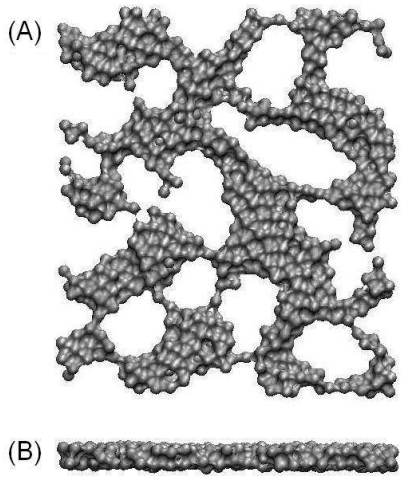}
  \caption{The structure of two consecutive Xe layers in UO$_{2}$ simulated at 1000\,K with an equal amount
  of xenon to uranium atoms on the same [111] layer. (A) top view and (B) side view.
}
  \label{fig:Xe-layer}
\end{figure}

From a point of topology, the layer structure of Xe on [111] plane
of UO$_{2}$ is kept very well. Figure \ref{fig:Xe-layer} gives
the final equilibrium configuration of [0.5{\_}0.5{\_}d] after a 15\,\emph{ps} simulation
in a supercell of $12\times12\times6$ (about $8.2\times9.5\times5.8nm^{3}$). It is interesting that all
Xe atoms connected together to form a huge net-like cluster expanding across [111] plane, as the top view shows.
In fact, except the case of [0.25{\_}0.25{\_}d] with the smallest number density of Xe
where isolated small Xe clusters are obtained, all other cases produce
a single large connected net-like cluster of Xe.
The side view in figure \ref{fig:Xe-layer}(B) demonstrates a perfect planar distribution of Xe in UO$_{2}$, and
it is surprising that no
any tendency towards bubble geometry was observed. Even increment temperature
to 3500\,K, quite similar results are still obtained.

\begin{figure}[!b]
\centering
  \includegraphics*[0.22in,0.19in][3.9in,2.96in]{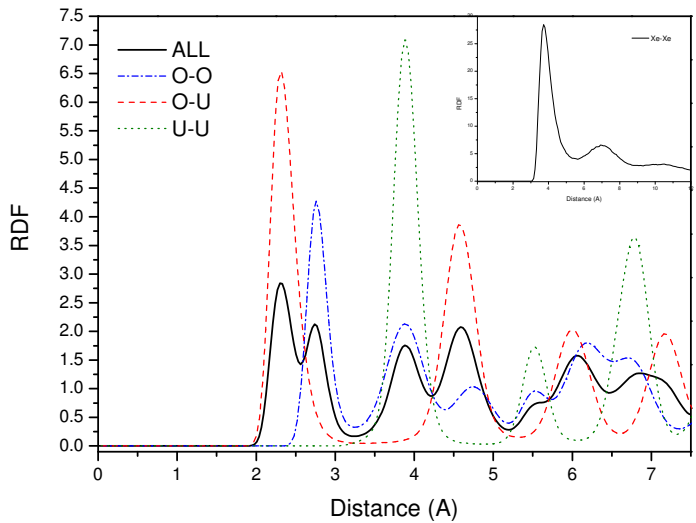}
  \caption{Radial distribution function (RDF) of UO$_{2}$ at 1000\,K of [0.5{\_}0.5{\_}d] configuration,
  where two consecutive [111] layers are doped
  with Xe by substituting half uranium atoms (as well as the bonded oxygen atoms to ensure the
  neutrality of simulation cell, see Fig.\ref{fig:fluorite}). The RDF among Xe atoms
  plotted in the inset shows typical characteristics of liquid.
}
  \label{fig:RDF}
\end{figure}

Since the attractive interaction between Xe atoms is very small, this
kind of aggregation evidently drives by uranium and oxygen atoms near the interface
which attempt to conserve the fluorite structure of UO$_{2}$ as perfect as possible.
The Xe-UO$_{2}$ boundary actually forms a solid-liquid interface here. The radial distribution
function (RDF) shown in figure \ref{fig:RDF} illustrates that Xe atoms are in a liquid state
while UO$_{2}$ matrix is in a solid state (see
the RDF of Xe-Xe pairs in the inset). It can be seen clearly from figure \ref{fig:RDF} that UO$_{2}$ is in a perfect solid state: the nearest neighbor (NN)
of O-U and O-O pairs comprise the first and second peaks of the total RDF,
the next nearest neighbor (NNN) O-O pair and NN pair of U-U, as well as Xe-Xe NN pair
contribute to the third peak. The total RDF of all atoms
also exhibits a characteristic of solid phase. But the inset graph gives an essential characteristic
of liquid for xenon itself. Therefore the behavior of Xe in UO$_{2}$ is quite
analogous to a system with hard sphere liquid in a solid.

\begin{figure}[!b]
\centering
  \includegraphics*[0.22in,0.19in][4.0in,2.96in]{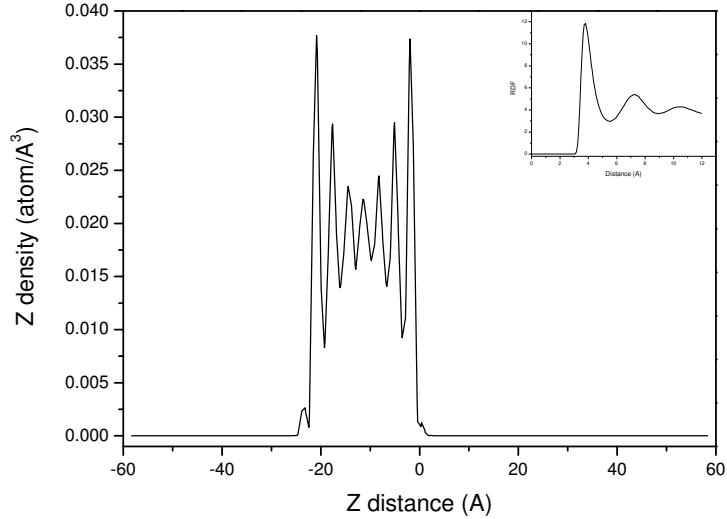}
  \caption{Variation of number density of Xe along (111) direction in
  UO$_{2}$ simulated at 1000\,K. A sandwich model is employed where initially six
  [111] layers are occupied exclusively by Xe atoms, other 30 layers are perfect UO$_{2}$.
  Inset shows the RDF of Xe-Xe pairs.
}
  \label{fig:Zden2}
\end{figure}

It is well known that near solid-liquid interfaces, for example a hard-sphere and
hard-wall systems or a crystal and its melt, the liquid phase will
present some oscillation of the density along the distance from the interface \cite{laird92,plischke86}.
This behavior results from the geometrical constraining effect of the solid surface
on the atoms or molecules of the liquid, which causes them to order into
quasi-discrete layers, similar to the oscillation of RDF of liquid. For liquid Xe having an interface with FCC aluminum,
similar behavior was observed \cite{donnelly02}. It is very analogous to what
was shown in figure \ref{fig:Zden1}, except here parts of solid fragment of UO$_{2}$ are mingled
with Xe liquid on the same layer.

\begin{figure}[!b]
\centering
  \includegraphics*[0.22in,0.19in][3.9in,2.96in]{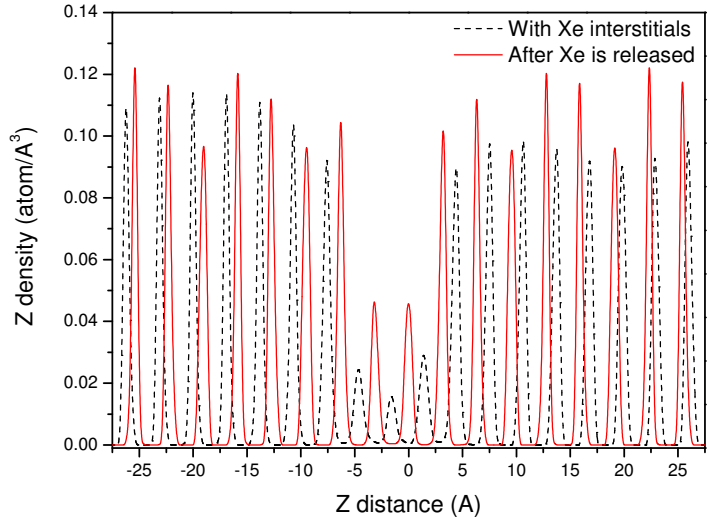}
  \caption{Recovery of the layer structure of uranium atoms along (111) direction after Xe
  interstitials are released out from UO$_{2}$ matrix, which initially has a distribution
  of xenon on [111] planes as [0.5{\_}0.5{\_}d], the same as figure \ref{fig:Xe-layer}.
}
  \label{fig:Zden-recov}
\end{figure}

It would be very interesting to study a case with a large amount of Xe atoms in UO$_{2}$,
where some layers are occupied by xenon exclusively. We made a
sandwich model that initially six consecutive [111] planes of FCC Xe are inserted into UO$_{2}$ in
a supercell of $6\times6\times12$ (about $4.1\times4.8\times11.7nm^{3}$), to
simulate the behavior of Xe near a solid-liquid interface. The variation of number
density along Z axis (perpendicular to the interface) is shown in figure \ref{fig:Zden2},
which predicts a layer-like (or planar) behavior of liquid Xe clearly. The layer structure
fades away along distance from the interface, but it still maintains beyond the fourth layer.
We also found that when UO$_{2}$ cannot preserve a good crystal structure, the
planar structure of Xe becomes blurry. Different from previous studies that mainly focused
on geometry constrained arised from solid surface, our simulation suggests that the planar distribution
of Xe atoms depends on internal state of solid phase directly (interaction among atoms)
and indirectly (structure relaxation and surface restructure).
The most interesting is that although different number density of xenon will leads to
different interval between layers, the interval distance of liquid Xe always equals to
that of UO$_{2}$.

\begin{figure}[!b]
\centering
  \includegraphics*[scale=0.5]{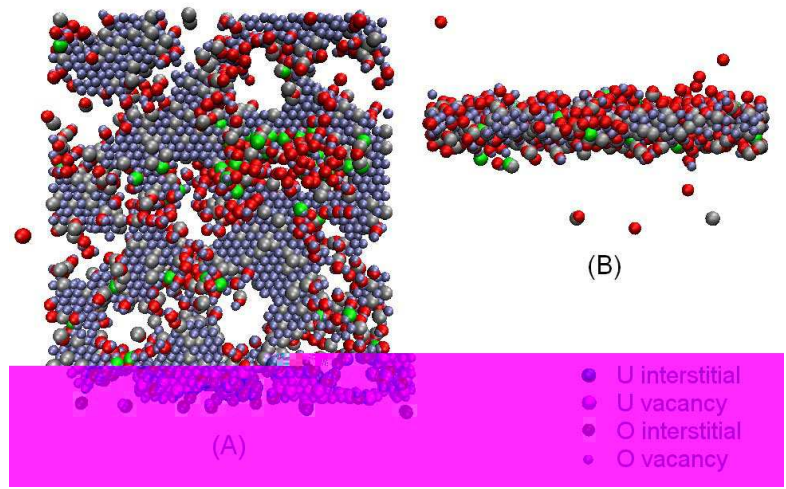}
  \caption{(Color online) The residual defects in UO$_{2}$ at 1000\,K after Xe
  (see figure \ref{fig:Xe-layer}) was released out: (A) is top view and (B) side view.
}
  \label{fig:24-24-defect}
\end{figure}

It is difficult to completely recover the damaged structure after Xe is released out at about 1000\,K.
This temperature is far below the melting point and cannot provide enough kinetic energy to overcome
recrystallization barriers. It is quite different from the situation
under irradiations, where excess energy from fission tracks can
melt or vaporize UO$_{2}$ locally and transport enough energy to matrix
to conquer the barriers. Figure \ref{fig:Zden-recov} compares the layer structure
along (111) direction of uranium before and after Xe interstitials are released out.
In spite of in a solid state, uranium atoms within the defected region still rearrange
the structure from 3 layers into 2 layers. However it cannot recover completely
and many vacancies and interstitials still remained, as figure \ref{fig:24-24-defect} shown.
Far from the defects left by released Xe atoms, UO$_{2}$ keeps
a perfect fluorite structure except few oxygen interstitials. The residual defects
loop inherits the planar geometry of foregoing Xe interstitials, partly due to the
mechanical stability of solid UO$_{2}$. The most remnant defects are uranium
and oxygen vacancies. It is reasonable because a plenty of UO$_{2}$ molecules have
been repulsed out of the simulation cell from their positions when fission gas Xe is
introduced. To eliminate the open space left by them, melting and annealing the
material is necessary.

\section{Conclusion}
\label{sec-Conclusion}

The transferability of classical potentials of widely used shell model
and pair potential model beyond fluorite phase was investigated by comparing
their predicted energy variations along compression and stretching of various structures
with \emph{ab initio} results. Some virtual structures
and intermediate phases along a structural transition were involved. Results
showed that despite shell model gives a precise description of fluorite phase of
UO$_{2}$, its applicability to other phases is questionable. The performance of
pair potential model suggested that the fault might originate from the classical treatment of charge
polarization. It arises a problem
for shell model on how to control the balance between precision and transferability.

Using pair potential model, we studied the planar clustering of Xe aggregating in UO$_{2}$.
Providing a quasi-layer distribution of xenon initially, our simulations showed it would be stable
even after the melting of UO$_{2}$. The Xe interstitials in UO$_{2}$ are in a
liquid state and have a layer
modulation of density perpendicular to [111] interface, which (1) has a tendency to connect all Xe atoms
together, except the case with a low Xe density, no isolated Xe cluster was observed; (2) has almost the same layer
distance as the solid part of UO$_{2}$; (3) maintains beyond four layers at least.
After releasing these Xe out, a planar defects loop left in the UO$_{2}$ matrix.
We believe it has some effects on the formation of rim structure of nuclear fuel materials.
In addition, it is necessary to point out that in this paper we did not consider the planar defects
perpendicular to other direction (say, (001) or (110)) that we prefer to leave for
later studies, where a similar conclusion can be expected.

\ack{}
This study was financially supported by the Budget for
Nuclear Research of the Ministry of Education, Culture, Sports,
Science and Technology of Japan, based on the screening and counseling by the
Atomic Energy Commission.

\end{document}